\documentclass{amsart}

\usepackage{amssymb}
\usepackage{amsfonts}
\usepackage{amsmath}
\usepackage{graphicx}%
\usepackage{lastpage}
\usepackage[legalpaper,bookmarks=true,colorlinks=true,linkcolor=blue,citecolor=blue]{hyperref}
\usepackage{fancyhdr}
\usepackage{color}
\usepackage[mathlines]{lineno}
\usepackage{lscape}
\usepackage{epsfig}
\usepackage{natbib}
\usepackage{float}

\theoremstyle{plain}

\numberwithin{equation}{section}

\begin{document}
\title{Measuring inequality: application of semi-parametric methods to real life data}
\author{ Tchilabalo Abozou Kpanzou $^{(1)}$}
\author{Tertius de Wet $^{(2)}$}
\author{G. S. Lo $^{(3,4,5)}$}
\email{nkpanzout@gmail.com, tdewet@sun.ac.za, gane-samb.lo@ugb.edu.sn}

\maketitle

\begin{abstract}
 A number of methods have been introduced in order to measure the inequality in various situations such as income and expenditure. In order to curry out statistical inference, one often needs to estimate the available measures of inequality. Many estimators are available in the literature, the most used ones being the non parametric estimators. \cite{kpanzou2011} has developed semi-parametric estimators for measures of inequality and showed that these are very appropriate especially for heavy tailed distributions. In this paper we apply such semi-parametric methods to a practical data set and show how they compare to the non parametric estimators. A guidance is also given on the choice of parametric distributions to fit in the tails of the data.\medskip
 
\noindent 
$^{1}$ Universit\'e de Kara, Togo (Tchilabalo Atozou Kpanzou)\\
$^{2}$Faculty of Economics and Management Science, University of Stellenbosch, Stellenbosch, South Africa\\
$^{3}$ LERSTAD, Gaston Berger University, Saint-Louis, S\'en\'egal (Gane Samb Lo, Modou Ngom). \\
$^{4}$ Affiliated to LSTA, Universit\'e Pierre et Marie Curie, Paris, France (Gane Samb LO)\\
$^{5}$ Associated to African University of Sciences and Technology, AUST, Abuja, Nigeria (Gane Samb Lo)\\
\newline

\noindent \textit{Corresponding author}. Tchilabalo Abozou Kpanzou. Email :
nkpanzout@gmail.com.\\

\noindent 
\textbf{keywords and phrases} : Income distribution; Inequality measures; Confidence intervals; Extreme Value Theory.\\
\noindent {\textbf{AMS 2010 Classification subjects}}. 62F10; 62G05; 62P05
\end{abstract}

\Large

\section{Introduction}
An area of application of the statistical methods is that of measures of inequality. These are very popular in economics and have applications in many other branches of Science, see e.g. \cite{eusilc2004}; \cite{hullingerschoch2009}; \cite{allison1978}. In order to curry out statistical inference, one often needs to estimate the available measures of inequality, namely the Gini, the Generalised entropy, the Atkinson, the quintile share ratio, just to mention a few. Many estimators are available in the literature, the most used ones being the non parametric estimators. Details on the definitions of inequality measures as well as non parametric estimators can be found in, e.g., \cite{cowellflach2007}, \cite{langeltille2011}, \cite{kpanzou2011}, \cite{kpanzou2014} and \cite{kpanzou2015}. Further on inequality and poverty measures can be found in \cite{lo2013} and \cite{loandmergane2013}.\\

\noindent \cite{kpanzou2011} has developed semi-parametric estimators for measures of inequality and showed that these are very appropriate especially for heavy tailed distributions. In this paper we apply such semi-parametric methods to a practical data set and show how they compare to the non parametric estimators. A guidance is also given on the choice of parametric distributions to fit in the tails of the data. The data used are claims data from a South African short term insurer.\\

\noindent The remainder of the paper is organized as follows. In Section \ref{section2}, we briefly describe the semi-parametric methods used. Results on application to the aforementioned data set are given in Section \ref{section3}. We give some concluding remarks in Section \ref{section4}.

\section{Methodology}\label{methodology}\label{section2}
In this section we present the semi-parametric estimation method. The procedure relies on the estimation of the underlying distribution in a semi-parametric setting.\\

\noindent Define a semi-parametric distribution function by
\begin{equation}\label{defspdistrfunct} \tilde{F}(x)=\begin{cases} & F(x), \ \ \ \ \ \ \ \ \ \ \ \ \ \ \ \ \ \ \ \ \ \ \ \ \ \ \ \ \ x\leq x_0,\\
	& F(x_0)+(1-F(x_0))F_{\theta}(x), \ x> x_0,
	
\end{cases}
\end{equation}

\noindent for a given $x_0$, where $F_\theta$ is a parametric distribution satisfying the condition $F_\theta(x_0)=0$, and $F$ is an unknown distribution. Note that $\theta$ can be a vector parameter. 
Choose $x_0=Q(F,1-\alpha), \ \alpha\in [0,1]$, where $Q$ denotes the quantile function associated with $F$.\\

\noindent We then have
\begin{align*}
F(x_0)+(1-F(x_0))F_{\theta}(x)&=1-\alpha+(1-(1-\alpha))F_\theta(x)\\
&=1-\alpha+\alpha F_\theta(x)\\
&=1-\alpha(1-F_\theta(x)).
\end{align*}

\noindent It follows that
\begin{equation}\label{semipardistrq} \tilde{F}(x)=\begin{cases}& F(x), \ \ \ \ \ \ \ \ \ \ \ \ \ \ \ \ \ x\leq Q(F,1-\alpha),\\
	& 1-\alpha(1-F_\theta(x)), \ x> Q(F,1-\alpha),
\end{cases}
\end{equation}

\noindent where $F_{\theta}$ satisfies the condition $F_\theta(Q(F,1-\alpha))=0$.\\

\noindent Estimating $\theta$ by $\widehat{\theta}$, and estimating $F$ by the empirical distribution function $F_n$, we estimate the underlying distribution semi-parametrically as
\begin{equation}\label{semipardistrqest} \tilde{F}_n(x)=\begin{cases}& F_n(x), \ \ \ \ \ \ \ \ \ \ \ \ \ \ \ \ x\leq Q(F_n,1-\alpha),\\
	& 1-\alpha(1-F_{\widehat{\theta}}(x)), \ x> Q(F_n,1-\alpha).
\end{cases}\end{equation}

\noindent Equation (\ref{semipardistrqest}) is very important as it estimates the underlying distribution. However, a choice of the parametric distribution $F_\theta$ is required in order to make the estimation process possible. We address that issue by making use of results from Extreme Value Theory (EVT). See e.g. Beirlant et al. \cite{beirlantetal2004} for these results.\\

\noindent Given a certain threshold $u$, we consider the conditional distribution of the exceedance of $u$ given that $u$ was exceeded. We consider two types of exceedances:
\begin{enumerate}
	\item $X-u$ given $X > u$ (absolute exceedance).
	\item $X/u$ given $X > u$ (relative exceedance).
\end{enumerate}

\noindent From EVT, if $F$ belongs to the domain of attraction of the generalized extreme value distribution, the following limiting results hold when $u\rightarrow\infty$.
\begin{enumerate}
	\item\label{evtres1} The distribution of $X-u|X > u$ converges to the generalized Pareto distribution (GPD)

\begin{equation}\label{gpddistrfct} 
	G(x;\sigma,\gamma)=1-\left(1+\frac{\gamma x}{\sigma}\right)^{-\frac{1}{\gamma}}, \ x>0,
\end{equation}

where $\sigma>0$ is the scale parameter and $\gamma>0$ is the extreme value index (EVI).
	\item\label{evtres2} The distribution of $X/u|X > u$ converges to the strict Pareto (Pa) distribution
	\begin{equation}\label{strpardistrfct} F_P(x)=1-x^{-\frac{1}{\gamma}}, \ x>1, \ \gamma>0,\end{equation}
	where $\gamma$ is the EVI.
	\item A second order approximation of the distribution of the relative exceedance is the perturbed Pareto distribution (PPD) defined by the survival function
	\begin{equation} \label{ppddistrfct} \overline{G}(x;\gamma,c,\tau) = 1-G(x;\gamma,c,\tau)=(1-c)x^{-1/\gamma}+cx^{-1/\gamma-\tau},\end{equation}
	where $x>1$, $\gamma>0$, $\tau>0$ and $c\in(-1/\tau,1)$. The idea here is to fit such a PPD to the relative exceedance, aiming for a more accurate estimation of the unknown tail. See \cite{beirlantetal2004} for more details.
	\end{enumerate}

\noindent Once the semi-parametric estimators for the distribution function are obtained, we plug them in the functional form of the inequality measures to obtain their SP estimators. Here we only give the estimator for Gini when fitting the GPD in order to illustrate the procedures we are aiming to apply.\\

\noindent Recall that given a distribution function $F$ of a random variable $X$ with mean $\mu$, the ordinary Gini coefficient can be defined as

\begin{equation} 
I_G=\frac{1}{\mu}\int_0^\infty F(x)(1-F(x))dx.
\end{equation}

\noindent Consider a random sample $X_1,X_2,\ldots,X_n$ from $F$, with associated order statistics $X_{1,n}\leq X_{2,n}\leq \ldots \leq X_{n,n}$, and suppose the threshold above which a parametric distribution is fitted is $x_0=Q(F,1-\alpha)$. The Gini coefficient is then estimated semi-parametrically as

\begin{equation} 
\widehat{I}_{SPG}=\frac{1}{\widehat{\mu}}\int_0^\infty \tilde{F}_n(x)(1-\tilde{F}_n(x))dx,
\end{equation}

\noindent where $\tilde{F}_n$ is given in Equation (\ref{semipardistrqest}), and $\widehat{\mu}$ is the estimator of $\mu$ using $\tilde{F}_n$, that is

\begin{equation}
\widehat{\mu}=\int_0^\infty xd\tilde{F}_n(x).
\end{equation}

\noindent Estimate the threshold $u$ by $X_{n-k,n}$, and assume that for a large $n$, the GPD is a reasonable approximation to the distribution of the exceedances $X_{n-k+1,n}-X_{n-k,n},X_{n-k+2,n}-X_{n-k,n},\ldots,X_{n,n}-X_{n-k,n}$ for a given $k$. A semi-parametric estimator for $I_G$ is then given by
\
\begin{equation} 
\widehat{I}_{SPG}=\frac{1}{\widehat{\mu}}\sum_{i=1}^{n-k-1}\frac{i}{n}(1-\frac{i}{n})(X_{i+1,n}-X_{i,n})+\frac{k\widehat{\sigma}\left[2n-k-\widehat{\gamma}(n-k)\right]}{n^2\widehat{\mu}(1-\widehat{\gamma})(2-\widehat{\gamma})},
\end{equation} 

\noindent where $\widehat{\sigma}$ and $\widehat{\gamma}$ are estimators for the unknown scale and shape parameters $\sigma$ and $\gamma$ of the GPD using the exceedances, with $\widehat{\gamma}<1$, and 
\begin{equation}
\widehat{\mu}=\frac{1}{n}\sum_{i=1}^{n-k}X_{i,n}+\frac{k}{n}\left(X_{n-k,n}+\frac{\widehat{\sigma}}{1-\widehat{\gamma}}\right)
\end{equation} 

\noindent is an estimator for $\mu$.\\

\noindent Theoretical details and other estimators can be found in \cite{kpanzou2011}. We remind that the aim of this work is to show how the semi-parametric methods can be applied to real life data, and so we direct the reader to the previous reference for all the theoretical derivations.\\

\section{Numerical applications and interpretations}\label{section3}
Here we consider claims data from a South African short term insurer. These consist of a portfolio of claims from 1 July 2004 to 21 July 2006. The dates used, were the dates the claims occurred, and not the dates the claims were registered. The claim amounts were the total claim amounts and any excesses paid by the client were ignored. The claim amounts were adjusted for inflation to July 2006 as base month. Finally, any negative or zero claim amounts were deleted from the data set. Negative amounts occur if, for instance, the value of the items salvaged from the wreck exceeds the claim amount in value. The final sample size is 16104. We will refer to the data set as Portfolio.\\

\begin{figure}
	\centering
	\includegraphics[width=8cm,height=6cm]{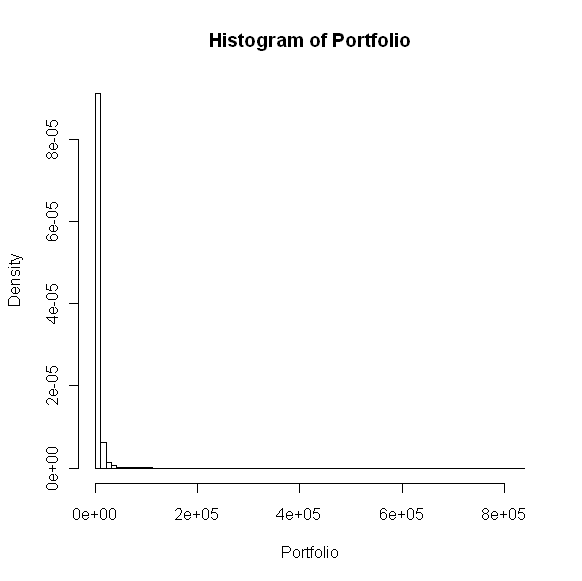}\\
	\includegraphics[width=8cm,height=6cm]{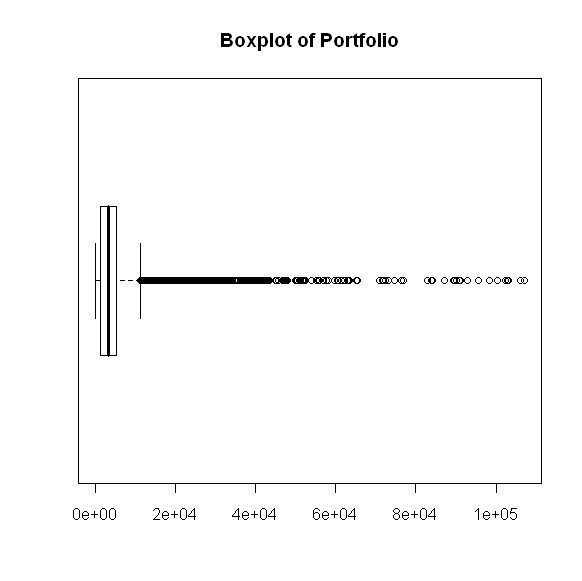}
	\caption{Histogram and Boxplot for Portfolio data}
	\label{histofnorpor123}
\end{figure}

\noindent The histogram and the boxplot for the data are shown in Fig. \ref{histofnorpor123}., giving an idea of the tails of the distribution.
\clearpage

\noindent We see from Fig. \ref{histofnorpor123}. that the data have heavy tails and so we can use them to illustrate our methods. Table \ref{sumarstatsappl} gives some descriptive statistics (Median, Median Absolute Deviation (MAD) and maximum value).
\begin{table}
\centering
\caption{Descriptive statistics for the Portfolio data set}
\begin{tabular}{|c|c|c|c|c|}
	\hline 
	& Sample size $n$ & Median & MAD & Maximum \\ 
	\hline 
	Portfolio & 16104 & 3268.5050 & 2951.3525 & 835567.7000 \\ 
	\hline 	
\end{tabular} 
\label{sumarstatsappl}
\end{table}

\noindent We next give the estimates of the inequality measures, starting with the non parametric ones summarised in Table \ref{nonparineqm}.
\begin{table}
\centering
	\caption{Non parametric estimates of inequality measures for Portfolio data set}
	\begin{tabular}{|c|c|c|c|c|}
		\hline 
		& Gini & GE0  & A1 & QSR \\ 
		\hline 
		Portfolio & 0.5767 & 0.7070  & 0.5069 & 34.8292 \\ 
		\hline 
	\end{tabular}
\label{nonparineqm}
\end{table}

\noindent The semi-parametric estimates for the inequality measures using the methods described in Section \ref{methodology} (fitting the GPD, the strict Pareto and the PPD in the tails) are now calculated. We use $10\%$ of the data (upper order statistics) in each case both to estimate the parameters in the tail distribution and to fit that distribution. For comparison purposes we put together both the non parametric and semi-parametric estimates. The results are given in Table \ref{allestimatesport}. We only show these for the Gini coefficient, the generalised entropy with parameter $0$ (GE0), the Atkinson with parameter $1$ (A1) and the quintile share ratio (QSR). The non parametric estimators are denoted by NP, the semi-parametric (SP) ones by SPGPD (when fitting the GPD in the tails), SPPa (when fitting the Pa) and SPPPD (when fitting the PPD).
\begin{table}
\centering
	\caption{Non parametric and semi-parametric estimates of inequality measures for Portfolio data set}
	\begin{tabular}{|c|c|c|c|c|}
		\hline 
		Portfolio & NP & SPGPD & SPPa & SPPPD \\ 
		\hline 
		Gini & 0.5767 & 0.5923 & 0.5145 & 0.5850 \\ 
		\hline 
		GE0 & 0.7070 & 0.7499 & 0.6999 & 0.7453 \\ 
		\hline 
		A1 & 0.5069 & 0.5276 & 0.5034 & 0.5072 \\ 
		\hline 
		QSR & 34.8292 & 36.6990 & 34.4200 & 34.9122 \\ 
		\hline 
	\end{tabular} 
	\label{allestimatesport}
\end{table}

\noindent We see from these tables that the SP estimates are not far away from the NP estimates. The interpretation is that for the portfolio data considered, we have a high level of inequality in the distribution. In fact, Gini is around $60\%$, GE0 around $70\%$, A1 around $50\%$, and the QSR shows that the $20\%$ upper claims are worth 35 times the $20\%$ lowers claims. For the interpretation of inequality measures, see e.g. \cite{creedy2014}.\\

\noindent However, the advantage of using the semi-parametric estimators relies on the fact that they are more resistant to extreme values in the data, as shown by \cite{kpanzou2011}. In addition, although we do not show confidence intervals in this paper, note that the coverage probabilities are very satisfactory for confidence intervals based on SP methods. This is also demonstrated in \cite{kpanzou2011}.\\

\noindent The question that is now raised is how one can choose the right parametric distribution to fit in the tails. Such a choice is done among the GPD, the Pa and the PPD. In order to decide which one to use in a specific application, we suggest the use of a measure of representativeness of the sample to each distribution. In our applications, we use the one given by \cite{bertino2006}. Below is the description of that measure.\\

\noindent Let $\underline{X}=(X_1,X_2,\ldots,X_n)$ be a simple random sample of size $n$ from a distribution $F$ and denote by $X_{1,n}\leq X_{2,n}\leq\ldots\leq X_{n,n}$ its associated order statistics. A representativeness index is given by

\begin{equation} 
R(\underline{X},F)=1-\frac{12n}{4n^2-1}\sum_{i=1}^n\left(F(X_{i.n})-\frac{2i-1}{2n}\right)^2.
\end{equation}

\noindent This index is used to measure how well the sample $\underline{X}$ represents its parent distribution. Therefore, it can be used to select the model one should use when having to choose between a number of different models.\\

\noindent In our case, we use the measure to decide between the three parametric choices, GPD, strict Pareto and PPD. To do this we calculate $R(\underline{X},GPD)$, $R(\underline{X},Pa)$ and $R(\underline{X},PPD)$, and the largest value determines the preferred distribution to use.\\

\noindent Applying the measure of representativeness to $10\%$ of the upper values of the Portfolio data considered, we obtain the results in Table \ref{measofreprstness}.
\begin{table}
\centering
	\caption{Measure of representativeness}
	\begin{tabular}{|c|c|c|c|}
		\hline 
		$\underline{X}$ & $R(\underline{X},GPD)$ & $R(\underline{X},Pa)$ & $R(\underline{X},PPD)$ \\ 
		\hline 
		Portfolio & 0.97362 & 0.99987 & 0.99988 \\ 
		\hline 
	\end{tabular} 
	\label{measofreprstness}
\end{table}

\noindent These results show that the strict Pareto and the PPD are well represented by the data in the tails and so either of them can be used. The corresponding estimators are the most reliable, and should be considered for assessing the inequality in the given situation.\\

\noindent Given a practical data set, in order to use a semi-parametric method, a two-step approach would be to first determine which of the three parametric distributions to use in the tail estimation. The corresponding semi-parametric procedure can then be used as a preferred choice to estimate the desired measures.

\section{Conclusions}\label{section4}
In this paper we have illustrated the application of semi-parametric estimators to real life data. Such estimators indeed perform very well, especially in the case of heavy tailed distributions. They are based on semi-parametric estimators of the underlying distribution using results from Extreme Value Theory. As the name indicates, a part of a semi-parametric estimator is made up with non parametric estimation and the other part (the tails) with parametric estimation. Three options have been considered for fitting such parametric distributions, namely the generalized Pareto, the strict Pareto and the perturbed Pareto. Given a data set, one therefore needs to decide which one of the three he should use for the estimation. We have addressed this question by suggesting the use of the measure of representativeness proposed by \cite{bertino2006}. In order to illustrate the way one should use it, this measure has also been applied to the portfolio data set considered.\\

\noindent From the analysis done in this paper, we make the following recommendations. Suppose one has a data set from a heavy tailed distribution. Then:

\begin{itemize}
	
\item[(i)] Choose a threshold above which a particular distribution fits well (we have the three possibilities mentioned above);
\item[(ii)] Use the measure of representativeness to choose the best fitting distribution;
\item[(iii)] Calculate the estimates of the inequality measures using the estimators corresponding to the distribution obtained in (ii).
\end{itemize}
Applying this procedure guarantees, to some extent, that the conclusions drawn from the analysis will be more reliable.

\end{document}